\documentclass[a4paper,12pt]{article}
\usepackage{amsfonts}
\usepackage{graphicx}
\usepackage{amsmath}
\usepackage{color}

\newcommand{\be}{\begin{equation}}
\newcommand{\ee}{\end{equation}}
\begin{document}


\begin{center}
\noindent{{\bf \LARGE{$Q$-curvature and gravity}}}

\smallskip
\smallskip

\smallskip
\smallskip

\smallskip
\smallskip

\smallskip
\smallskip

\noindent{{Mariano Chernicoff$^a$, Gaston Giribet$^{b}$, Nicol\'as Grandi$^c$, Edmundo Lavia$^{d,e}$,  Julio Oliva$^f$}}

\smallskip
\smallskip

\smallskip
\smallskip

\smallskip
\smallskip

\smallskip
\smallskip

\smallskip
\smallskip

$^a${Departamento de F\'{\i}sica, Facultad de Ciencias, Universidad Nacional Aut\'onoma de M\'exico}\\ 
{\it A.P. 70-542, CDMX 04510, M\'exico.}

\smallskip

$^b${Center for Cosmology and Particle Physics, New York University}\\
{\it 726 Broadway 10003 New York City, USA.}

\smallskip

$^c${Instituto de F\'{\i}sica de La Plata - CONICET \& Departamento de F\'{\i}sica - UNLP}\\ 
{\it C.C. 67, 1900 La Plata, Argentina.}

\smallskip

$^d${Departamento de F\'{\i}sica, Universidad de Buenos Aires \& IFIBA - CONICET}\\ 
{\it Ciudad Universitaria, pabell\'on 1 (1428) Buenos Aires, Argentina.}

\smallskip

$^e${Argentinian Navy Research Office (DIIV), UNIDEF and CONICET \\
{\it Laprida 555, 1638 Vicente L\'opez, Buenos Aires, Argentina.}}

\smallskip

$^f${Departamento de F\'{i}sica, Universidad de Concepci\'on}\\ 
{\it Casilla 160-C, Concepci\'on, Chile.}


\end{center}

\smallskip
\smallskip

\smallskip
\smallskip

\smallskip
\smallskip

\smallskip

\begin{center}
{\bf Abstract}
\end{center}
In this paper, we consider a family of $n$-dimensional, higher-curvature theories of gravity whose action is given by a series of dimensionally extended conformal invariants. The latter correspond to higher-order generalizations of the Branson $Q$-curvature, which is an important notion of conformal geometry that has been recently considered in physics in different contexts. The family of theories we study here includes special cases of conformal invariant theories in even dimensions. We study different aspects of these theories and their relation to other higher-curvature theories present in the literature.



\newpage

\section{Introduction}

Quantum effects induce higher-curvature modification to the gravitational action. This is well understood in the context 
of string theory, where the ultraviolet corrections to the low energy effective action can be systematically computed 
\cite{Gross:1986iv}. On general grounds, higher-curvature modifications render the theory of gravity renormalizable, but 
at the cost of introducing ghost instabilities \cite{Stelle:1976gc} and other pathologies \cite{Jose, Gruzinov, Hofman}. This
implies that, whatever higher-curvature correction to Einstein theory to be proposed, it has to satisfy very special constraints in order to be physically acceptable
\cite{Zwiebach}. One may still ask whether such constraints are restrictive enough to define the theory uniquely or, on 
the contrary, there exist more than one consistent way of modifying general relativity (GR). In fact, there 
are known higher-curvature actions that define theories with interesting properties and which, under certain conditions, 
no longer have ghosts. 

One such example is the so-called Critical Gravity\footnote{See the discussion in \cite{Porrati} and references therein.} (CG), which is defined by supplementing the 
Einstein-Hilbert action on Anti-de Sitter (AdS) space with a conformally invariant linear combination of $R^2$ terms 
with a
specific value of the coupling constant \cite{lu.2011}. The precise linear combination corresponds to the square of 
the Weyl tensor, i.e. $L^2\int d^4x\sqrt{-g} \: C_{\alpha \beta \mu \nu}C^{\alpha \beta \mu \nu}$, where  the coupling 
constant $L^2$, having mass dimension $-2$, is adjusted in terms of the cosmological constant $\Lambda $. In dimension 
$n=4$, the theory includes general relativity (GR) as a particular subsector, is free of the massive spin-$0$ mode that 
quadratic theories typically engender, and acquires a second massless spin-$2$ mode apart from the GR graviton. The 
presence of a second massless spin-$2$ field produces low-decaying modes and it causes the black holes and other 
solutions of the theory to have vanishing gravitational energy. 

Critical Gravity theories can also be defined in higher dimension, $n> 4$ \cite{deser.2001}. This amounts to 
dimensionally continue the $4$-dimensional conformal invariant by simply replacing the action with $L^2\int d^nx\sqrt{-g} 
\: C_{\alpha \beta \mu \nu}C^{\alpha \beta \mu \nu}$ and chose the coupling constants in such a way that the maximally 
symmetric vacuum is unique. As in $4$ dimensions, CG in $n>4$ has no massive modes; the spin-$0$ conformal mode 
decouples and the extra spin-$2$ mode becomes massless. However, in contrast to $n=4$, in dimension $n>4$ CG does not 
generically admit Einstein spaces as solutions; the reason being the presence of the Kretschmann scalar 
$R_{\mu\nu\rho\sigma}R^{\mu\nu\rho\sigma}$ in the action, which in $n>4$ contributes dynamically. This does not happen 
for $n=4$ in virtue of the Chern-Weil-Gauss-Bonnet theorem \cite{Lanczos}. The latter represents the main difference 
between CG in $n=4$ and $n>4$.        

Another higher-curvature theory that exhibits special features is Lovelock theory \cite{Lovelock:1972vz, Lovelock:1971yv}, which is defined by dimensionally extending topological invariants to higher $n$. The resulting theory coincides with GR only in dimension $n\leq 4$, while in $n>5$ presents higher-curvature corrections up to order $R^{k}$, with $k<n/2$. Despite involving contractions of more than one Riemann tensor in the Lagrangian, Lovelock action yields second-order field equations. In fact, Lovelock field equations are the most general covariantly conserved symmetric rank-2 tensor in dimension $n$ that is of second order in the metric and torsion free. For $n=4$ the latter requirements single out the Einstein tensor, while in $n\geq 5$ they allow for more tensor structures. Lovelock field equations, however, contain higher powers of the second derivatives of the metric, unlike GR. This makes the dynamical structure of the theory to exhibit special features that give rise to peculiar physical phenomena \cite{Gomberoff}.

Here, we will investigate a class of higher-curvature theories which are different from CG and Lovelock theories but 
nonetheless share some features with both of them. In fact, the family of theories we propose to explore can be thought 
of as a hybrid between CG and Lovelock models, in the sense that {are} defined by dimensionally 
extending conformal invariants, in opposition to topological invariants. In dimension $4$, these theories include 
conformal gravity and CG as particular cases. In dimension greater than $4$, in contrast,
they do
not agree with the $n$-dimensional generalization of \cite{deser.2001} and 
they
can rather be regarded as a different way of extending the CG  of \cite{lu.2011} to arbitrary $n$. 
They do include,
nevertheless, 
other higher-dimensional theories  recently considered in the literature; in particular, for $n=6$ 
they include 
the cubic theories studied in 
reference \cite{Lu:2011ks}. 

Other differences with CG and Lovelock theories are the following: Unlike Lovelock theory, the one we propose to study here modifies GR even for $n\leq 4$. On the other hand, unlike the $n>4$ CG theories of \cite{deser.2001}, our theory does admit generic Einstein spaces as solutions. The price to be paid is that the spin-$0$ massive excitation around AdS$_n$ does not decouple and dealing with this requires further imagination. There exists, however, a choice of coupling constant that renders the extra spin-$2$ mode massless. In addition to Einstein spaces, which persist as solutions up to a renormalization of the cosmological constant, the theory also admits non-Einstein solutions, as we will see. 

The fundamental building block to construct the action of the theory will be the so-called $Q$-curvature, which is an important notion of conformal geometry \cite{GrahamJuhl, Juhl}. Originally introduced by Branson in \cite{B}, the $Q$-curvature is a local scalar quantity that plays an important role in topics as diverse as spectral geometry, conformal geometry, differential topology and the theory of higher-order differential equations, among others. Recently, $Q$-curvature has also been studied in theoretical physics; in particular, to study anomalies in quantum field theory \cite{N}, higher-derivative field theories \cite{Oz}, and other related problems. In section 2, we will review the definition and the main properties of the $Q$-curvature, together with its higher-dimensional and higher-order generalizations. In section 3, we will discuss its connection to conformal invariants in even dimensions. This will provide us with the ingredients to construct, in section 4, the gravitational action of our theory. In section 5, we will discuss the simplest solutions of the theory: their maximally symmetric 
vacua.
We will derive the conditions to have a unique such vacuum and for the linear excitations around 
it
to become massless. Section 6 contains comments about the black hole solutions, the expressions of their charges and the associated thermodynamics variables. In section 7, we will explore the non-linear gravitational wave solutions. Non-Einstein spaces will be discussed in section 8, where we will provide explicit examples in dimension $n=5$. These examples include black holes, product of spherical spaces and their squashed deformations, and AdS$_2\times M$ solutions. In section 9, we will comment on other higher-curvature actions also associated to the $Q$-curvature.
We will comment on the relation between these theories and other models such as New Massive Gravity, Critical Gravity, and the counterterms that appear in the context of holographic renormalization.

\section{$Q$-curvature}

In order to introduce the notion of $Q$-curvature and motivate its definition, we will begin by revisiting properties of higher-curvature terms under conformal transformations: 
Given the Weyl rescaling 
of an $n$-dimensional metric 
\begin{equation}
g_{\mu\nu}\to \tilde{g}_{\mu\nu}=e^{2\varphi}g_{\mu\nu}
\end{equation}
we
consider a linear differential operator $P_{m,n}$ with $m\in 2\mathbb{Z}_{\geq 0}$, $n\in \mathbb{Z}_{\geq 0}$ that transforms covariantly as follows
\begin{equation}
\tilde{P}_{m,n}(f)=e^{-\frac{n+m}{2}\varphi}P_{m,n}(e^{\frac{n-m}{2}\varphi}f),
\end{equation}
with $P_{0,n}:=1$. Here, $f$ represents an arbitrary differentiable function. In other words, $\tilde{P}_{m,n}$ is an $m^{\text{th}}$-order linear differential operator of conformal bi-degree $(\frac{n-m}{2},\frac{n+m}{2})$. This operator $P_{m,n}$ has the form
\begin{equation}
P_{m,n}=\Delta_{m,n}+\frac{n-m}{2}Q_{m,n} \ , \ \ \ \ \ \Delta_{m,n} = \Box^{\frac m2} + ...\label{ElP}
\end{equation}
with $\Box=g^{\mu\nu}\nabla_{\mu}\nabla_{\nu}$ being the Laplace-Beltrami operator. The ellipsis stand for terms with no constant term, i.e. $\Delta_{m,n}$ is a linear differential 
operator satisfying $\Delta_{m,n}1 =0$. $Q_{m,n}$ is a scalar curvature that transforms as follows
\begin{equation}
\tilde{Q}_{m,n}= e^{-\frac{n+m}{2}\varphi} \Big( Q_{m,n}+\frac{2}{n-m}\Delta_{m,n}\Big)e^{\frac{n-m}{2}\varphi},\label{LaQ}
\end{equation}
and is what is called the $m^{\text{th}}$-order, $n$-dimensional $Q$-curvature, which satisfies $(n-m)Q_{m,n} = 2P_{m,n}(1)$. 

The transformation laws above uniquely define the linear operators $P_{m,n}$ and the scalars $Q_{m,n}$. The simplest example of the hierarchy (\ref{ElP})-(\ref{LaQ}) (i.e. $m=2$) is
\begin{equation}
Q_{2,n}=-\frac{1}{2(n-1)}R \ , \ \ \ P_{2,n}=\Box +\frac{n-2}{2}Q_{2,n}  \ , \ \ \ \Delta_{2,n}=\Box .
\end{equation}
That is, $Q_{2,n}$ corresponds to the Gaussian curvature and $P_{2,n}$ to the Yamabe operator
\begin{equation}
P_{2,n}=\Box -\frac{n-2}{4(n-1)}R.
\end{equation}

Branson's $Q$-curvature corresponds to the case $m=4$, which takes the form
\begin{equation}
Q_{4,n} = -\frac{1}{2(n-1)} \Box R -\frac{2}{(n-2)^2} R_{\mu \nu}R^{\mu \nu} + \frac{n^2(n-4)+16(n-1)}{8(n-1)^2(n-2)^2} R^2,
\end{equation}
where $P_{4,n}$ is the so-called Paneitz operator; see (\ref{Panes}) below. Operator $P_{4,n}$ was originally defined by Fradkin and Tseytlin in \cite{FT} and independently by Riegert in \cite{R}. 

The case $m=6$ takes the form
\begin{eqnarray}
Q_{6,n} &=&-\frac{1}{32\left(  n-4\right)  \left(  n-2\right)  ^{2}\left(  n-1\right)
^{3}} \Big( \left(  n^{5}-8n^{4}+64n^{3}-240n^{2}+1008n-960\right)
R^{3}+ \nonumber \\
&&\ \ \ \ \ 512\left(  n-1\right)
^{3}R^{\mu \nu }\square R_{\mu \nu }-4\left(  n-1\right)  \left(  n^{4}-14n^{3}%
+100n^{2}-168n+96\right)  R\square R-\nonumber \\
&& \ \ \ \ \ 64\left(  n-1\right)  ^{2}\left(  n^{2}-4n+28\right)
RR_{\mu \nu }R^{\mu \nu }+
1024\left(  n-1\right)  ^{3}R_{\alpha \beta}R_{\mu \nu }R^{\alpha \mu \beta \nu }\Big).\label{Q66}
\end{eqnarray}
In $n=6$ and up to boundary terms, (\ref{Q66}) coincides with the particular combination of conformal invariants proposed in \cite{Lu:2011ks}, which has the property of being the unique conformal invariant combination in 6 dimensions that admits generic Einstein manifolds as solutions. This provides us with a criterion to select our theory and define the general Lagrangian of order $m$, in dimension $n$: We will consider Lagrangians consisting of dimensionally extended conformal invariants and that preserve Einstein spaces as solutions. 

The hierarchy $Q_{m,n}$ continues ad infinitum, although the expressions become cumbersome for $m>6$. The case $m=8$, 
for example, is a dimension 8 operator involving quartic operators such as $R^4$, $R^2R_{\mu \nu } R^{\mu \nu }$, 
$(R_{\mu \nu }R^{\mu \nu })^2$, $RR_{\mu \alpha \nu \beta}R^{\alpha \beta } R^{\mu \nu }$, ... $R_{\mu\nu} \Box^2 R^{\mu\nu}$, $R \Box^2 R$, whose explicit form can be found 
in \cite{Gover}. Written in terms of the Schouten tensor $P_{\mu \nu }=(R_{\mu \nu}-Rg_{\mu\nu}/(2n-2))/(n-2)$ and the 
Weyl tensor {
$C_{\mu\nu\alpha\beta} = R_{\mu\nu\alpha\beta} + g_{\alpha\nu}P_{\mu\beta} - g_{\alpha\mu}P_{\nu\beta} 
+ g_{\beta\mu}P_{\nu\alpha} - g_{\beta\nu}P_{\mu\alpha} $,
}
the expression for $Q_{8,n}$ simplifies notably, but the number of terms 
still rises to more than forty.

\section{Conformal invariants}

Now, let us comment on the connection between $Q$-curvature and conformal invariants. We begin by reviewing well-known facts of 2-dimensional manifolds: Consider a closed Riemann surface with Euclidean signature $(M_2,g)$. According to the Gauss-Bonnet theorem, its Euler characteristic, $\chi (M_2)$, is computed by the integral 
\begin{equation}
\mathcal{I} =-\frac{1}{2\pi}\int_{M_2}d^2x\sqrt{g} \: Q_{2,2} =\frac{1}{4\pi}\int_{M_2}d^2x\sqrt{g}R = \chi(M_2) 
\label{ERTYU}
\end{equation}
where $g$ is the determinant of the Euclidean metric $g_{\mu\nu}$, and $R$ is the Ricci scalar (i.e. the Gaussian curvature). This is a topological invariant. In dimension $2$, all metrics are locally conformally equivalent and we also have the following properties: Provided one rescales the metric as $g_{\mu \nu }\to  e^{2\varphi }{g}_{\mu \nu }$ the Ricci scalar transforms as $ R\to e^{-2\varphi }(R-2\Delta_{2,2} \varphi)$ while the Laplace-Beltrami operator transforms simply as $ \Delta_{2,2} \to e^{-2\varphi }\Delta_{2,2} $. These transformations are important to understand in what sense the Branson $Q$-curvature is the natural generalization of Gauss curvature to dimension $4$. To motivate the definition of the $Q$-curvature \cite{B, BO}, let us explicitly write the Paneitz operator \cite{P},
\begin{equation}
P_{4,4}=\Delta_{4,4} = (\Box )^2+2G_{\mu \nu }\nabla^{\mu }\nabla^{\nu }+\frac 13 (\nabla^{\mu }R_{\mu \nu})\nabla^{\nu }+\frac 13 R\Box .\label{Panes}
\end{equation}
where $G_{\mu \nu }=R_{\mu \nu }-(1/2)Rg_{\mu \nu}$ is the Einstein tensor. This is a linear fourth-order, four-dimensional differential operator that under the rescaling of the metric $g_{\mu \nu }\to  e^{2\varphi }{g}_{\mu \nu }$ transforms as $\Delta_{4,4} \to e^{-4\varphi }\Delta_{4,4} $. From this, the definition of the $Q$-curvature is natural: It is the fourth-order, four-dimensional curvature invariant that, having the same scaling dimension than $\Delta_{4,4} $, transforms simply as $Q_{4,4} \to e^{-4\varphi } (\Delta_{4,4} \varphi +Q_{4,4}) $. This has the form
\begin{equation}
Q:=Q_{4,4}= -\frac 16 \Box R -\frac 12 R_{\mu \nu }R^{\mu \nu } +\frac 16 R^2. 
\end{equation}

To reinforce the analogy with what Gaussian curvature $R\propto Q_{2,2}$ means in dimension $n=2$, let us mention that in the same way as how $Q_{2,2}$ computes the Euler characteristic in $2$ dimensions, $Q_{4,4}$ computes the Euler characteristic $\chi(M_4)$ of a 4-dimensional Riemann manifold $(M_4,g)$ within a particular conformal class. More precisely,
\begin{equation}
\mathcal{I}=\frac{1}{8\pi^2}\int_{M_4}d^4x \sqrt{g} \: Q_{4,4} +\frac{1}{32\pi^2} \int_{M_4}d^4x \sqrt{g} \: C_{\mu \nu 
\alpha \beta}C^{\mu \nu \alpha \beta} = \chi(M_4)\label{Ultima}
\end{equation}
where $C_{\mu \ \alpha \beta}^{\ \nu}$ is the Weyl tensor. Notice that both terms on the left hand side are conformal invariants. That is, $Q$-curvature computes a topological invariant within a given conformal class. In dimension 2, of course, there is only one conformal class and thus (\ref{Ultima}) turns out to be a natural generalization of (\ref{ERTYU}).

Branson also provided \cite{B} a definition of the $Q$-curvature in arbitrary dimension $n>3$. For $n\neq 4$, its definition is given in terms of its transformation rules under Weyl rescaling and not by its topological meaning. This is given by
\begin{equation}
Q_{4,n}= A_n \Box R +B_n R_{\mu \nu}R^{\mu \nu} + C_n R^2,\label{Q}
\end{equation} 
with $A_n =-{1}/({2(n-1)})$, $B_n=-{2}/{(n-2)^2}$, $C_n=(n^2(n-4)+16(n-1))/({8(n-1)^2(n-2)^2})$.

This is the second term in the list of scalars $Q_{m,n}$ we discussed in the previous section. In particular, all the 
integrals $\int d^nx\sqrt{-g}\: Q_{n,n}$ are conformal invariants. The scalars $Q_{m,n}$ will constitute the 
Lagrangian density of the theory we propose to explore.

\section{The action}

The gravity action we will consider is defined by the sum of the dimensionally continued conformal invariants; namely
\begin{equation}
\mathcal{I} = \int d^nx\sqrt{-g}\ \sum_{k=0}^{\infty} L^{2k-2}b_{k}P_{2k,n}(1)\label{quince} 
\end{equation}
where $P_{2k,n}(1)=(n/2-k)Q_{2k,n}$, with $k\in \mathbb{Z}_{\geq 0}$, and where $P_{0,n}=1=(n/2)Q_{0,n}$. We are now considering $n$-dimensional pseudo-Riemannian manifold $(M_n,g)$ with Lorentzian mostly plus signature. $L$ is a constant of mass dimension $-1$. This sets the length scale $L$ at which the ultraviolet corrections due to the higher-curvature terms $Q_{m>2,n}$ start to contribute significantly. The dimensionless coupling constants $b_k$ are usually normalized in such a way that $b_0=-\Lambda L^2/(8\pi G)$ and $b_1=-(n-1)/(4\pi G (n-2))$, where $G$ is the $n$-dimensional Newton constant. Our conventions will be such that $b_2=-1/(4\pi G (n-4)^2)$. That is,
\begin{equation}
\mathcal{I} = \frac{1}{16\pi G}\int_{M_n} d^nx\sqrt{-g} \Big( R-2\Lambda + \frac{4L^2}{(n-2)^2(n-4)} \Big( R_{\mu \nu }R^{\mu \nu }- \frac{n^3-4n^2+16n-16}{16(n-1)^2}R^2\Big)+ ...\Big) ,
\end{equation}
where the ellipsis stand for higher-curvature, higher-derivative terms. 

Of course, for $b_{k>1}=0$ action (\ref{quince}) reduces to Einstein theory. Other particular choices are also interesting: The case $b_k=\delta_{2,k}$ for $n=4$ corresponds to 4-dimensional conformal gravity. The special case $b_0=-\Lambda L^2/(8\pi G)$, $b_1=-3/(8\pi G)$, $b_2=-L^2/(4\pi G (n-4))$ with $L^2=3/(2\Lambda )$ in the limit $n\to 4$ reduces to the Critical Gravity theory proposed in \cite{lu.2011}; see also \cite{Miskovic:2014zja}. The case $b_k=\delta_{3,k}$ for $n=6$ corresponds to the cubic theory defined in \cite{Lu:2011ks}, whose action is given by the linear combination of conformal invariants in $6$ dimensions that supports Einstein manifolds as solutions. In general, action (\ref{quince}) with $b_{k}=\delta_{n/2,k}$ defines a conformal invariant theory, classically.  

The theory described by (\ref{quince}) with $b_k=\delta_{2,k}$ in arbitrary dimension $n$ is also special: Defined on a 
closed Euclidean $n$-dimensional manifold $(M_n,g)$, it corresponds to the variational problem of minimizing the Branson 
$Q$-curvature on $M_{n}$. For $n>4$, the Euler-Lagrange equations derived from such action,
{$E_{\mu \nu }:=\delta\mathcal{I}/\delta g^{\mu \nu}=0$,}
have trace equal to $Q_{4,n}$. (Therefore, turning on $b_0\neq 
0$ yields field equations whose solutions solve the uniformization problem $Q_{4,n}=const$ on $M_n$). For 
$b_k=\delta_{2,k}$ in dimension $n>4$, the tensor $E_{\mu \nu }$ obeys the following three properties: $E:=g^{\mu \nu 
}E_{\mu \nu} =  Q_{4,n}$, $E_{\mu \nu} = E_{\nu \mu}$, and $\nabla^{\mu}E_{\mu \nu} = 0$. That is, it is a covariantly 
conserved, symmetric rank-2 tensor whose trace is the $Q$-curvature. These properties are reminiscent of the properties 
that Lin and Yuan required to define their $J$-tensor in \cite{LY}, i.e. a symmetric rank-2 tensor canonically 
associated to the $Q$-curvature. However, the divergence of the $J$-tensor does not vanish but it turns out to be 
proportional to the gradient of $Q$. More precisely, the Lin-Yuan $J$-tensor obeys: $ J:=g^{\mu \nu }J_{\mu \nu} =  
Q_{4,n}$, $ J_{\mu \nu} = J_{\nu \mu}$, and $\nabla^{\mu}J_{\mu \nu} = ({1}/{4})\nabla^{\mu}Q_{4,n}$. The motivation to 
define such a tensor is the following: If one insists with the idea that $Q$-curvature is the fourth-order analogue of 
the Gaussian curvature $R$, then a natural question is what is the analogue of the Ricci tensor $R_{\mu \nu}$ and of its 
derived notions such as Ricci-flatness, Einstein manifolds, etc. To answer this question, one recalls the basic 
properties of $R$ and $R_{\mu \nu }$, namely: $g^{\mu \nu }R_{\mu \nu} =  R$, $R_{\mu \nu} = R_{\nu \mu}$, and 
$\nabla^{\mu}R_{\mu \nu} = ({1}/{2})\nabla^{\mu}R$. Then, the analogy becomes evident: In the same manner as how the 
$Q$-curvature can be regarded as the fourth-order generalization of $R$, the tensor $J_{\mu \nu }$ turns out to be the 
generalization of the Ricci tensor $R_{\mu \nu }$. From this, definitions such as $J$-flatness, $J$-Einstein, etc follow 
naturally. Along the same lines, our tensor $E_{\mu \nu }$ should be regarded as the natural fourth-order generalization 
of Einstein tensor $G_{\mu\nu}$, and thus it is natural to consider it as the completion of our gravity field equations. 
The precise relation between our tensor $E_{\mu \nu }$ and the Lin-Yuan tensor is
\begin{equation}
E_{\mu \nu} =\frac{4}{(4-n)}\Big( J_{\mu \nu}-\frac 14 g_{\mu \nu} J \Big) \ , \ \ \ \ \ 
J_{\mu \nu} =\frac{(4-n)}{4}E_{\mu \nu}+\frac 14 g_{\mu \nu} E ;
\end{equation} 
with $J=E=Q_{4,n}$. Summarizing, our action (\ref{quince}) provides a definition of the Einstein-Hilbert variational problem for the Lin-Yuan $J$-tensor, i.e. it gives an action functional definition of $J_{\mu \nu}$ (for $n>4$). 

\section{Vacua}

Now, we go back to the interpretation of action (\ref{quince}) as defining a theory of gravity. For concreteness, we focus on the case that includes higher-curvature terms up to the quadratic order $Q_{m\leq 4,n}$. In this case, the action is given by
\begin{equation}
\mathcal{I} = \frac{1}{16\pi G}\int d^nx \sqrt{-g} \Big( R-2\Lambda + \alpha R^2 + \beta R_{\mu \nu }R^{\mu \nu }\Big)
\end{equation}
with
\begin{equation}
\alpha = -L^2 \frac{(n^3-4n^2+16n-16)}{4(n-1)^2(n-2)^2(n-4)} \ , \ \ \ \ \ \ \beta =L^2\frac{4}{(n-2)^2(n-4)}.
\end{equation}

This theory admits solutions of constant curvature, namely 
\begin{equation}
R_{\mu \alpha \nu \beta} = -\frac{1}{\ell^2} (g_{\mu \nu }g_{\alpha \beta}- g_{\mu \beta }g_{\alpha \nu})
\end{equation}
which are maximally symmetric spaces obeying the Einstein equations 
\begin{equation}
R_{\mu \nu } = -\frac{(n-1)}{\ell^2} g_{\mu \nu }\label{P21}
\end{equation}
with a curvature radius $\ell$ given by
\begin{equation}
{\Lambda \ell^4}+\frac{(n-1)(n-2)\ell^2}{2}+\frac{(n+2)(n-2)L^2}{8}=0.\label{veintidos}
\end{equation}

This equation, for $n>4$, yields two values for $\ell^2$. Generically, the theories with $Q_{2k,n}$ contains $k$ maximally symmetric vacua with different curvature radii. For special choices of the coupling constants $b_k$, however, some of these vacua degenerate. For instance, the condition for (\ref{veintidos}) to yield a unique vacuum reads
\begin{equation}
L^2=-2\ell^2\frac{(n-1)}{(n+2)}.
\end{equation}
In this case, the theory has a unique maximally symmetric solution with an effective cosmological constant 
{$\Lambda_{\text{eff}}=-(n-2)(n-1)/(4\ell^2)$.}
The condition for this unique vacuum to be AdS$_n$ is $\ell^2 >0$, i.e. 
$L^2 <0$, $\alpha >0$, $\beta <0$. 

For arbitrary $\ell^2/L^2$, the degrees of freedom of fluctuations about AdS$_n$ 
include a massless spin-$2$ mode, and a massive spin-$0$ mode. These modes are typically tachyonic. In fact, demanding 
the effective Newton constant to be positive one finds that one of the two spin-$2$ fields has a mass $m^2_{s=2} = 
-(n-2)^2( (n^2-4)+2({\ell^2}/{L^2})(n-1)(n-4)) / (8\ell^2(n-1))$; (hereafter $16\pi G =1$, unless explicitly declared). 
One can easily choose the value of the coupling constant $L^2$ such that $m^2_{s=2}=0$. In that case, as we will see, 
also the black hole solutions of the theory become massless. The massive spin-$0$ mode, on the other hand, has mass $ m^2_{s=0} = (n-1)( 4m^2_{s=2}-({2}/{L^2})(n-2)^2)/(n-2)^2$. One can in principle accept the values $m^2_{s}<0$ and compare them with the Breitenlohner-Freedman (BF) bound: $m^2_{s} \geq m^2_{\text{BF}}=- ((n-1)^2+4s)/(4\ell^2)$. This posses a bound for $L^2$, which is $n$ dependent. The scalar conformal mode is frequently the most problematic. We will discuss in section 9 a series of theories that permits to decouple this mode. There exist different 
ways of dealing with it: One way is considering values of the coupling constant such that the mass of this mode becomes 
infinite and it eventually decouples \cite{Bueno1, Bueno3, Bueno4, Bueno5}. Another possibility is to 
look for boundary conditions that suffice to eliminate the mode in a dynamically consistent way, cf. \cite{Lu:2011ks, 
Lu:2011mw, Maldacena:2011mk} . One could also investigate special type of matter to which the theory can be coupled 
without the scalar mode to introduce pathologies. Another logical possibility is invoking non-linear effects that cure 
the theory. Last, one can also look for other backgrounds 
around 
 which the modes result well defined.

\section{Black holes}

Theory (\ref{quince}) admits Einstein spaces (\ref{P21}) as solutions, provided $\ell$ satisfies (\ref{veintidos}). In particular, it contains black holes. The metric of a AdS-Schwarzschild black hole is given by
\begin{equation}
ds^2 = -\Big( 1-\frac{r_0^{n-3}}{r^{n-3}}+\frac{r^2}{\ell^2}\Big) dt^2 + \Big( 1-\frac{r_0^{n-3}}{r^{n-3}}+\frac{r^2}{\ell^2}\Big)^{-1} dr^2 + r^2 d\Omega_{n-2}^2 \label{IO89}
\end{equation}
where $d\Omega^2_{n-2}$ is the metric on the unit $(n-2)$-sphere and $r_0$ is an integration constant associated to the mass. In fact, the mass of this black hole solution is given by \cite{Deser:2007vs, Deser:2005jf, Deser:2002jk, Deser:2002rt, Olea}
\begin{equation}
M_{\text{BH}}= \frac{1}{8\pi G}\Big( 1+\frac{L^2(n-2)(n+2)}{2\ell^2(n-1)(n-4)}\Big) (n-2)\text{Vol}(\Omega_{n-2})r_0^{n-3} \label{T28}
\end{equation}
where we have reinserted the overall normalization $(16\pi G)^{-1}$ in the action. $\text{Vol}(\Omega_{n-2})$ in (\ref{T28}) stands for the volume of the $(n-2)$-sphere, namely $\text{Vol}(\Omega_{n-2}) = {2\pi^{\frac{n-1}{2}}}/{\Gamma(\frac{n-1}{2})}.$

The Hawking temperature associated to the black hole solution (\ref{IO89}) is
\begin{equation}
T_{\text{H}} =  \frac{(n-1)r_+^2+(n-3)\ell^2}{4\pi \ell^2r_+},
\end{equation}
which is a geometrical quantity and consequently independent of the presence of higher-curvature terms. In contrast, the entropy does depend on the coupling constant $L$ in a way that can be computed by different methods. The result reads
\begin{equation}
S_{\text{BH}} = \frac{\text{Vol}(\Omega_{n-2})r_+^{n-2}}{4G}\Big( 1+\frac{L^2(n-2)(n+2)}{2\ell^2(n-1)(n-4)}\Big) = \frac{\text{Area}}{4G} + \mathcal{O}(L^2/\ell^2) 
\end{equation}
where the first term between brackets gives the Bekenstein-Hawking contribution $\text{Area}/(4G)$, accompanied by higher-curvature corrections to the prefactor. Notice that the entropy $S_{\text{BH}}$ and the mass $M_{\text{BH}}$ satisfy the first principle $dM_{\text{BH}}=T_{\text{H}}dS_{\text{BH}}$. It is also easy to check that both $S_{\text{BH}}$ and $M_{\text{BH}}$ vanish when the mass of the spin-2 fluctuating mode, $m^2_{s=2}$ is zero.

\section{Gravitational waves}

Now, we move to explore exact gravitational wave solutions. We consider the ansatz 
\begin{equation}
d{s}^{2}=\frac{\ell^{2}}{r^{2}}\left( -\left( 1+2{H} \right) dt^{2}+2dtd\xi
+dr^{2}+\delta_{ij}dx^idx^j\right) ,  \label{wave}
\end{equation}%
where ${H} $ is a function that does not depend on the lightlike
coordinate $\xi $. Here, $\delta_{ij}$ is the $(n-3)$-dimensional Kronecker delta that defines the Euclidean metric on
$\mathbb{R}^{n-3}$. We consider deformations of the universal
covering of AdS$_{n}$, so the coordinates take values $t\in \mathbb{%
R}$, $\xi \in \mathbb{R}$, and $r\in \mathbb{R}_{\geq 0}$. ${H} =const$ corresponds to AdS$_{n}$ space in Poincar\'{e} coordinates, with its
boundary located at $r=0$. For the deformation, we consider the null geodesic
vector $k^{\mu }\partial _{\mu }=(r/l)\partial _{\xi }$, which enables to
interpret these backgrounds as Kerr-Schild transformations of AdS$_{n}$; namely 
\begin{equation}
g_{\mu \nu }=g_{\mu \nu }^{\mathrm{AdS}}-2{H} \ k_{\mu }k_{\nu }.
\label{eq:K-S}
\end{equation}%
where $g_{\mu \nu }^{\mathrm{AdS}}$ is the metric of AdS$_{n}$; recall $k_{\mu
}k^{\mu }=0$.

The Ricci tensor for a metric like (\ref{eq:K-S}) takes the form 
\begin{equation}
R_{\mu \nu }=-\frac{(n-1)}{\ell^{2}}g_{\mu \nu }+k_{\mu }k_{\nu }\square {{H} }%
,  \label{eq:Ricci}
\end{equation}%
and it yields constant scalar curvature $R=-n(n-1)/\ell^{2}$, which turns out to be independent of $H$. It also yields the dimension $6$ operators
\begin{eqnarray}
R_{\mu \alpha }R_{\nu }^{~\alpha } &=&\frac{(n-1)^{2}}{\ell^{4}}g_{\mu \nu }-%
\frac{2(n-1)}{\ell^{2}}k_{\mu }k_{\nu }\square {{H} }, \\
R_{\mu \alpha \nu \beta }R^{\alpha \beta } &=&\frac{(n-1)^{2}}{\ell^{4}}g_{\mu
\nu }-\frac{(n-2)}{\ell^{2}}k_{\mu }k_{\nu }\square {{H} },\qquad 
\\
R_{\mu \gamma \alpha \beta }R_{\nu }^{~\gamma \alpha \beta } &=&\frac{2(n-1)%
}{\ell^{4}}g_{\mu \nu }-\frac{4}{\ell^{2}}k_{\mu }k_{\nu }\square {{H} },
\end{eqnarray}
and 
\begin{eqnarray}
\square {R}_{\mu \nu }&=&k_{\mu }k_{\nu }\square \left( \square -\frac{2}{\ell^{2}
}\right) {H} .
\end{eqnarray}

Using the expression for the Ricci tensor and the properties of $k^{\mu }$, one finds that the only non-trivial contribution to the field equations is
\begin{equation}
\ k_{\mu } k_{\nu } \ (\square -M^{2})\square {H} =0.  \label{GGGG}
\end{equation}
with $M^{2}$ being given by
\begin{equation}
M^2=-\frac{(n-2)^2}{8\ell^2 (n-1)}\Big( (n^2-4) + 2\frac{\ell^2}{L^2} (n-1)(n-4)\Big). \label{Malados}
\end{equation}
The condition for (\ref{Malados}) to be zero is
\begin{equation}
\ell^2 = -L^2\frac{(n-2)(n+2)}{2(n-1)(n-4)},
\end{equation}
and we observe that when $M^2=0$ the gravitational energy of the AdS-Schwarzschild black hole is also zero. This is analogous to what happens in CG in arbitrary dimension \cite{waves}. Another special value for $M^2$ is the one for which the AdS$_n$ vacuum results unique. This happens when 
\begin{equation}
M^2_{0} = -\frac{(n-2)^2(n+2)}{4\ell^2 (n-1)}.
\end{equation}

\section{Non-Einstein spaces}

Besides Einstein-spaces, theory (\ref{quince}) admits a large class of non-Einstein solutions. Among them, there are solutions with anisotropic scale invariance, with and without Galilean symmetry. That is, the theory admits both Shr\"{o}dinger \cite{Ayon} and Lifshitz \cite{Ayon2} type metrics for specific values of the dynamical exponent, $z$. There is another class of solutions given by the direct product of squashed or stretched deformations of AdS spaces and constant curvature spaces. This class includes the so-called Warped-AdS$_3$ spaces, Warped-AdS$_3$ black holes, and AdS$_2 \times S^1$ spaces. To be concrete, let us focus on the $5$-dimensional case for which such metrics take the form 
\begin{equation}
	ds^2 = \frac{\ell^2}{\mu^2 + 3}\left( -\cosh^2(r)dt^2 + dr^2 + 
	\frac{4 \mu^2}{\mu^2 + 3}(dx + \sinh(r)dt )^2  + d\Sigma^2_{2,\pm } \right)\label{YUIO}
\end{equation}
where $d\Sigma^2_{2,\pm }$ is a metric of a 2-dimensional space of constant curvature $\pm 1$; namely
\begin{equation}
	d\Sigma^2_{2,+ } =\tau^2 ( dy^2 + \sin^2(y)dz^2) \ , \ \ \ \ d\Sigma^2_{2,- } = \tau^2 ( dy^2 + \cosh^2(y)dz^2)\ 
, 
\end{equation}
with $\tau^2$ being a constant that controls the radius of the internal 2-dimensional piece of the geometry, $\Sigma_{2,\pm}$. We can take $t\in \mathbb{R}$, $x\in \mathbb{R}$, and $r\in \mathbb{R}$. These coordinates parameterize the 3-dimensional part of the geometry that describes a squashed or stretched deformation of AdS$_3$, also known as Warped-AdS$_3$ spaces or simply WAdS$_3$. The parameter that controls the deformation is $\mu$;
the value 
 $\mu=1$ 
corresponding 
to the undeformed AdS$_3$ space written as a Hopf fibration of AdS$_2$. The scalar curvature associated to the 5-dimensional geometry (\ref{YUIO}) is 
\begin{equation}
	R = -\frac{2 \, {\left(3 \, \tau^{2} \mp  {\mu}^{2} \mp 3\right)}}{\tau^{2} {\ell}^{2}}
\end{equation}
where the squashing parameter $\mu $ is related to the radius $\tau$ by 
\begin{equation}
	\mu^2 =  {\frac{3(1\pm \tau^2)}{X_{\pm }(\tau)} \ , \ \ \  \text{with}} \ \ \ X_{\pm }(\tau) = 2\tau^4\pm 
	5\tau^2 -1 ,
\end{equation}
and where the coupling constants take the values
\begin{equation}
	L^2 = {48\ell^2}\frac{X_{\pm}(\tau )}{Y_{\pm}(\tau )} \ , \ \ \ \ 
	\Lambda = -\frac{3}{2\ell^2}\frac{Z_{\pm}(\tau)}{X_{\pm}(\tau)Y_{\pm}(\tau)}
\end{equation}
with
\begin{equation}
	Y_{\pm }(\tau) = 78\tau^4\mp 267 \tau^2 -145 \ , \ \ \
	Z_{\pm}(\tau) = 156\tau^8 \mp 556 \tau^6 -2661 \tau^4 \pm 666\tau^2 +1015
\end{equation}

Warped AdS$_3$ spaces admit black hole solutions \cite{Clement} that are asymptotically WAdS$_3$ as well as locally WAdS$_3$ \cite{Strominger}, and they also admit a limit in which the geometry becomes AdS$_2\times S^1$. All these spaces have very interesting properties and deserve to be studied separately. 

\section{Alternative dimensional extension}

There exists another way of dimensionally extending to $n\geq 4$ the theory that, in $n=4$, is defined by considering the sum of scalars $Q_{2k\leq 4,4}$ in the Lagrangian density. To see this, let us be reminded of the fact that in $4$ dimensions one has
\begin{equation}
Q_{4,4}+  \frac 14 C_{\mu \nu \alpha \beta}C^{\mu \nu \alpha \beta}= \frac 14 \mathcal{E}_4  -\frac{1}{6} \Box R \ ,\label{HU}
\end{equation} 
where the right hand side is a total derivative as it includes $\Box R$ and the Pfaffian $\mathcal{E}_4=R_{\mu \nu \alpha \beta}R_{\mu \nu \alpha \beta}$ $-4R_{\mu \nu}R_{\mu \nu }+R^2$. 

While Lovelock theory corresponds to dimensionally extending the right hand side of (\ref{HU}), the theory discussed in the preceding sections corresponds to extending the $Q$-curvature by replacing $Q_{4,4}$ by $Q_{4,n}$. However, this is not the only way in which one can extend (\ref{HU}) to $n>4$ dimensions as one could alternatively consider the combination $\mathcal{E}_4-C_{\mu \nu \alpha \beta}C^{\mu \nu \alpha \beta}$ and then extend both the Gauss-Bonnet term $\mathcal{E}_4$ and the Weyl tensor $C^{\ \nu}_{\mu \ \alpha \beta}$ to $n$ dimensions. To see that the latter differs from the simple extension $Q_{4,4}\to Q_{4,n}$, let us notice that in $n$ dimensions the following identity holds
\begin{equation}
Q_{4,n}+\frac {1}{4} C_{\mu \nu \alpha \beta}C^{\mu \nu \alpha \beta} - \frac{1}{4}\mathcal{E}_4 = -A_n\Box R + \hat{\alpha } R^2 + \hat{\beta } R_{\mu \nu}R^{\mu \nu} \label{OPIO}
\end{equation} 
where
\begin{equation}
\hat{\alpha }= -\frac{(n-4)(2n^3-5n^2+6n-4)}{8(n-1)^2(n-2)^2} \ , \ \ \ \ \ \hat{\beta }= \frac{(n-1)(n-4)}{(n-2)^2} .\label{HJK}
\end{equation}
We see from this that the right hand side of (\ref{OPIO}) is a total derivative only for $n=4$. Therefore, in $n>4$ 
there exist two possibilities to define a higher-curvature theory based on the dimensional extensions of identity 
(\ref{HU}); namely either one considers the action $\int d^nx\sqrt{-g} \: Q_{4,n}$, as we did in the preceding 
sections, or one considers the action $\int d^nx\sqrt{-g} \: (\mathcal{E}_4 - C_{\mu \nu \alpha \beta}C^{\mu \nu \alpha 
\beta})$. Let us now explore the latter possibility; namely, consider the Lagrangian density 
\begin{equation}
\mathcal{L}_2 = L^2\Big( \mathcal{E}_4 - C_{\mu \nu \alpha \beta}C^{\mu \nu \alpha \beta}\Big) = \frac{n(n-3)L^2}{(n-1)(n-2)}R^2 - \frac{4(n-3)L^2}{(n-2)}R_{\mu \nu }R^{\mu \nu },\label{supercritica}
\end{equation}
with a coupling constant $L^2$. This theory exhibits interesting properties. In fact, it can be alternatively defined by minimal requirements: the absence of the conformal mode $\Box R$, the persistence of Einstein manifolds as solutions, and the uniqueness of the maximally symmetric vacuum. To see this, let us introduce the notation $\mathcal{L}_2=\alpha R^2+\beta R_{\mu \nu}R^{\mu \nu}+\gamma  R_{\mu \nu \rho\eta}R^{\mu \nu \rho\eta}$ with coupling constants $\alpha$, $\beta$, $\gamma $. The requirement of Einstein spaces to persist as solutions demands the coupling constant of the Kretschmann scalar, $\gamma $, to be zero. Next, the condition of the conformal mode to decouple yields the relation 
\begin{equation}
\alpha = -\frac{n\beta }{4(n-1)},
\label{alph}
\end{equation}
which makes $\Box R$ to disappear from the trace of the field equations. This is exactly the value of the relative 
coefficient that appears in the counterterm expansion of the boundary action in holographic renormalization 
\cite{haro.2001,henningson.1998,balasu.1999,hyun.1999}. Also, related to that, (\ref{alph}) agrees with the relative 
coefficient of the action that governs the induced gravity on a co-dimension 1 surface in AdS$_n$ gravity 
\cite{myers.2013}. Equation (\ref{alph}) has also relation with theories in lower dimension: For $n = 2$, it corresponds 
to $\alpha /\beta =-1/2$, for which the quadratic terms disappear from the action. For $n = 3$, it yields $\alpha /\beta 
=-3/8$, which corresponds to the so-called New Massive Gravity (NMG) introduced in \cite{bergshoeff.2009}. For $n = 4$, 
(\ref{alph}) yields $\alpha /\beta =-1/3$, and the quadratic piece of the action is, up to a total derivative, the 
conformal invariant combination $C_{\mu\nu\rho\sigma}C^{\mu\nu\rho\sigma}$. The $n>4$ CG theory of \cite{deser.2001}, 
however, does not agree with (\ref{supercritica}), (\ref{alph}){, but actually}
corresponds to the values 
\begin{equation}
\alpha = -\frac{\beta }{2(n-1)}, \ \ \  \gamma = -\frac{(n-2)\beta }{4}, \ \ \  \text{with} \ \ \ \Lambda = -\frac{(n-1)}{2(n-3)\beta }.
\label{coefs2}
\end{equation}

Last, the condition for the maximally symmetric vacuum of the theory to be unique yields the relation
\begin{equation}
\Lambda = \frac{(n-1)}{2(n-4)\beta }
\label{Lamb}
\end{equation}
which is valid for $n \neq 4$. This implies that the effective curvature radius is given by
\begin{equation}
\ell^2 = -\frac{(n-2)(n-4)}{2}\:\beta .
\end{equation}
In $n=3$, for instance, this agrees with the special point $\ell^2=\beta/2$ at which NMG exhibits special features \cite{NMG2, JulioTempoTroncoso}. 

In summary, there exists an alternative quadratic theory of gravity for $n>4$ that is special and is originally motivated by extending the $4$-dimensional Lagrangian density $Q_{4,4}$ to higher dimensions. This is defined by the coefficients 
\begin{equation}
\alpha = -\frac{n\beta }{2(n-1)}, \qquad \gamma = 0, \qquad \Lambda = -\frac{(n-1)}{2(n-4)\beta }.
\label{coefs}
\end{equation}
cf. (\ref{coefs2}). This theory and, in particular its relation to holographic renormalization deserve further analysis.

\section*{Acknowledgments}

The authors are grateful to Prof. Andr\'es Anabal\'on and Universidad Adolfo Ib\'a\~nez at Vi\~na del Mar for the hospitality during the early stages of this work. 
G.G. thanks Eloy Ay\'on-Beato, Mokhtar Hassa\"{\i}ne, Olivera Mi\v{s}kovi\'{c}, Rodrigo Olea, and David Rivera for many interesting discussions on related matters. He also thanks Sema Salur for the invitation to present this work at the AMS meeting held in Northeastern University, in Boston, in April 2018. The work of G.G. is supported in part by the NSF through grant PHY-1214302.
This work was also partially supported by CONICYT grant PAI80160018 and Newton-Picarte Grants DPI20140053.
M.C. is partially supported by Mexico's National Council of Science and Technology (CONACyT) grant 238734 and DGAPA-UNAM grant IN113618. M.C. dedicates this work to Olivia Chernicoff.

\end{document}